# How the relativistic motion affects Einstein-Podolsky-Rosen steering

Wen-Yang Sun[1], Dong Wang[1, 2] and Liu Ye[1,*]

[1] *School of Physics & Material Science, Anhui University, Hefei 230601, People's Republic of China*

[2] *CAS Key Laboratory of Quantum Information, University of Science and Technology of China, Hefei 230026, People's Republic of China*

**Abstract:** In this Letter, the dynamic behavior of Einstein-Podolsky-Rosen (EPR) steering and the redistribution of EPR steering under a relativistic framework are investigated. Specifically, we explore the scenario that particle *A* hold by Alice is in a flat space-time and another particle *B* by Bob entangled with *A* is in a non-inertial framework. The results show that EPR steering from Alice to Bob is dramatically destroyed by Unruh effect caused by the acceleration of Bob. Besides, EPR steering has an asymmetry property, and EPR steering asymmetry increases with the growing intensity of Unruh effect, implying that the Unruh effect can bring on EPR steering asymmetry. Furthermore, the reduced physical accessible EPR steering from Alice to Bob is distributed to the physical inaccessible EPR steering (from Alice to anti-Bob or from Bob to anti-Bob). Notably, unlike entanglement and quantum discord, only one of EPR steering from Alice to anti-Bob and Bob to anti-Bob experiences a sudden birth with the increase of acceleration parameter, which means that they cannot simultaneously survive. That is, the monogamy relation of EPR steering is still tenable in such a scenario. Consequently, we believe that EPR steering could also be served as one of important information resources within the long-distance quantum secure communication under the relativistic framework.

## 1. Introduction

Quantum entanglement is defined as the nonseparability of quantum states [1-3], and is one of the most important resources in quantum information processing (QIP) [4-8]. It has been a topic of great interest ever since the pioneering work was presented by Einstein *et al.* [9] in 1935. Meanwhile, the phenomenon of the Einstein-Podolsky-Rosen (EPR) steering was introduced by Schrödinger [10, 11] in 1935 to analyze the EPR-paradox. Subsequently, several theoretical and experimental works concerning EPR steering have been achieved [12-23], and formulated steering in an operational way in conformity for a quantum information task by Wiseman, Jones, and Doherty [24]. Recently, EPR steering was given an operational explanation as the distribution of entanglement by an untrusted party [25, 26]. Besides, the EPR steering has one-way property [20], namely, one quantum state may be steerable from Alice to Bob, however Bob cannot steer Alice.

---

[*] **Corresponding author: yeliu@ahu.edu.cn**

Further, EPR steering is an intermediate type of quantum correlations between Bell nonlocality [1] and entanglement [6] in modern quantum information theory. Apart from the fundamental interest in EPR steering, there is also an application in quantum key distribution [27]. Importantly, EPR steering can be detected by violating EPR steering inequalities [28-31], the violation of which provides one sufficient condition for a given quantum state to be steerable. The significant EPR steering criteria has been developed [32-36] to test EPR steering from different fields. Additionally, these criterions can also be used to guarantee one-way steering [15], and EPR steering and its asymmetry have been verified in some theoretical and experimental works [12-20]. Furthermore, the monogamy of EPR steering inequality violation was introduced by Reid [37] in 2013, the monogamy relationship indicates that two systems *A* and *C* cannot both steer a third system *B*.

From another point of view, understanding quantum phenomena in a relativistic framework is of basic importance because the realistic quantum regimes are essentially non-inertial. There is an arisen area of study for QIP under a relativistic framework, both theoretical and experimental tasks [38-43]. Besides, the relativistic effects of the Earth obviously affect satellite-based QIP tasks [44] and clock synchronization [45]. It was generally believed that the relativistic effects can cause the degradation of entanglement shared between an inertial observer and an accelerated one [41-43]. Consequently, it is of great interest to investigate how relativistic effects influence the properties of EPR steering. As a matter of fact, there are few authors to pay attention to address this open problem [46].

In this task, there are two parties, sharing an entangled state, Alice and Bob. Assume that Alice and Bob initially share a two-mode parametrized state. Alice is an inertial observer who stays stationary at a flat space-time region, while Bob is a non-inertial observer traveling with a uniform acceleration *a*. Under the non-inertial framework, mode *B* is mapped into two sets of regions for the outside region I and the inside region II, respectively. The complete description of the system involves three modes: mode *A*, described by the observer Alice; mode $B_\text{I}$, described by the observer Bob of outside region; and mode $B_{II}$, described by a hypothetical observer anti-Bob confined inside region. Our aim is to investigate the dynamic behavior of EPR steering under the influence of Unruh effect [47] caused by Bob's acceleration, and derive the EPR steering $S^{A \to B_I}$, which quantifies to what degree Alice can steer Bob's state by her measurements, and EPR steering $S^{B_I \to A}$, to certify the asymmetric property of EPR steering with Unruh effect. In addition, the redistribution of EPR steering is also put forward in detail. The research results indicate that EPR steering has several interesting physical characteristics as follows: (i) the maximally entangled mixed states (EMSs) can lead to maximal steerability, and all EMSs can be employed to realize EPR steering from Alice to Bob when the state parameter $\alpha$ is approximately larger than 0.6 and less than $\pi/2$. (ii) The reduced physical accessible EPR steering from Alice to Bob is distributed to the physical inaccessible EPR steering (from Alice to anti-Bob or from Bob to anti-Bob). This means that Unruh effect can destroy the physical accessible EPR steering and enhance the physical inaccessible EPR steering. (iii) Unlike entanglement and quantum discord [48], EPR steering from Alice to anti-Bob and from Bob to anti-Bob cannot simultaneously arise caused by the monogamy relation of EPR steering [37]. And the EPR steering has asymmetry property under the relativistic framework as well.

The letter is structured as follows. In Sect. 2, the EPR steering of bipartite *X*-state is reviewed. In Sect. 3, the dynamic behavior of EPR steering in a relativistic framework is investigated. In Sect. 4, EPR steering redistribution is presented. In final, we summarize our paper in the last section.

## 2. EPR steering of bipartite *X*-state

It is well known that the bipartite *X*-state can be expressed as

$$\rho^X = \begin{pmatrix} \rho_{11} & 0 & 0 & \rho_{14} \\ 0 & \rho_{22} & \rho_{23} & 0 \\ 0 & \rho_{23} & \rho_{33} & 0 \\ \rho_{14} & 0 & 0 & \rho_{44} \end{pmatrix}, \quad (1)$$

where $\rho_{ij}(i,j=1,2,3,4)$ are all real parameters. Then, by employing appropriate local unitary transformations, the *X*-shaped state can be rewritten as

$$\rho^X = \frac{1}{4}\begin{pmatrix} 1+c_3+s+r & 0 & 0 & c_1-c_2 \\ 0 & 1-c_3+r-s & c_1+c_2 & 0 \\ 0 & c_1+c_2 & 1-c_3-r+s & 0 \\ c_1-c_2 & 0 & 0 & 1+c_3-r-s \end{pmatrix}, \quad (2)$$

with

$$c_1 = 2(\rho_{23}+\rho_{14}),\ c_2 = 2(\rho_{23}-\rho_{14}),\ c_3 = \rho_{11}-\rho_{22}-\rho_{33}+\rho_{44},$$
$$r = \rho_{11}+\rho_{22}-\rho_{33}-\rho_{44},\ s = \rho_{11}-\rho_{22}+\rho_{33}-\rho_{44}. \quad (3)$$

in Bloch decomposition.

Here we introduce the criterion of EPR steering by choosing entropic uncertainty relation (EUR) steering inequality. If the entangled states can violate the steering inequality, the states are steerable. According to the entropic steering inequality's definition in [31], by using discrete EUR $H_q(R)+H_q(S) \geq \log(\Omega)$ with $\Omega \equiv \min_{i,j}\left(1/|\langle R_i|S_j\rangle|^2\right)$, along with our local hidden state model constraint for discrete observables $H(R^B|R^A) \geq \sum_\lambda P(\lambda)H_q(R^B|\lambda)$, we can immediately arrive at a EUR steering inequality for pairs of discrete observables [31]

$$H(R^B|R^A) + H(S^B|S^A) \geq \log(\Omega^B), \quad (4)$$

where $\Omega^B$ is the value $\Omega$. One realizes that for any EUR, even some relating more than two observables, there is a corresponding steering inequality [31]. In addition, from Refs. [31, 33, 49], one sees that, for even $N$, the EPR steering inequality can be given by

$$\sum_k^{N+1} H\left(R_k^B|R_k^A\right) \geq G_{EVEN}, \quad (5)$$

where $H(B|A) = H(\rho_{AB}) - H(\rho_A)$ is the conditional von Neumann entropy, and $G_{EVEN} \equiv \sum_i^{N+1} H(R_i) \geq (N/2)\log(N/2) + (1+N/2)\log(1+N/2)$, here, $N$, the system's dimensionality, is a positive integer power of a prime number. Besides, for odd $N$, $G_{ODD} \equiv \sum_i^{N+1} H(R_i) \geq (1+N)\log[(1+N)/2]$. Herein, $G_{ODD}$ and $G_{EVEN}$ are defined as the bounds for these EUR to condense these expressions later. These EUR can be adapted into steering inequality readily by substituting conditional entropies for marginal ones. Note that here and throughout the paper the base of all logarithms is assumed to be 2.

In terms of the EPR steering inequality (5), in two-dimensionality system $N=2$, utilizing the Pauli $X$, $Y$, and $Z$ measurements, then, we can obtain EPR steering $A \to B$ iff the condition [31]

$$SI = H\left(\sigma_x^B \big| \sigma_x^A\right) + H\left(\sigma_y^B \big| \sigma_y^A\right) + H\left(\sigma_z^B \big| \sigma_z^A\right) \geq 2 \tag{6}$$

is violated. For conveniently, employing Eqs. (2) and (6), we can define the EPR steering $A \to B$ to quantify how much the bipartite $X$-state is steerable by using the Pauli $X$, $Y$, and $Z$ measurements

$$S^{A \to B} := \max\left\{0, \frac{SI^{A \to B} - 2}{SI_{\max} - 2}\right\}, \tag{7}$$

with

$$\begin{aligned}SI^{A \to B} = &\sum_{i=1,2}\left[(1+c_i)\log(1+c_i) + (1-c_i)\log(1-c_i)\right] - (1+r)\log(1+r) - (1-r)\log(1-r) \\ &+ \frac{1}{2}\big[(1+c_3+r+s)\log(1+c_3+r+s) + (1+c_3-r-s)\log(1+c_3-r-s) \\ &+ (1-c_3-r+s)\log(1-c_3-r+s) + (1-c_3+r-s)\log(1-c_3+r-s)\big] \leq 2,\end{aligned} \tag{8}$$

and $SI^{A \to B} \leq SI_{\max} = 6$ (when the state is a maximally entangled pure state, $SI_{\max} = 6$), one guarantees that $S \in [0, 1]$. In the same manner, the corresponding expression of EPR steering $B \to A$ can be obtained by exchanging the roles of $A$ and $B$ in Eq. (6), resulting in an expression as if Eq. (7)

$$S^{B \to A} := \max\left\{0, \frac{SI^{B \to A} - 2}{SI_{\max} - 2}\right\}, \tag{9}$$

with

$$\begin{aligned}SI^{B \to A} = &\sum_{i=1,2}\left[(1+c_i)\log(1+c_i) + (1-c_i)\log(1-c_i)\right] - (1+s)\log(1+s) - (1-s)\log(1-s) \\ &+ \frac{1}{2}\big[(1+c_3+r+s)\log(1+c_3+r+s) + (1+c_3-r-s)\log(1+c_3-r-s) \\ &+ (1-c_3-r+s)\log(1-c_3-r+s) + (1-c_3+r-s)\log(1-c_3+r-s)\big] \leq 2.\end{aligned} \tag{10}$$

If $r = s = 0$, the bipartite $X$-state will become the Bell-diagonal state. The Eq. (8) can be simplified into $SI^{A \to B} = \sum_{i=1}^{3}\left[(1+c_i)\log(1+c_i) + (1-c_i)\log(1-c_i)\right] \leq 2$ [33].

## 3. Analyzing the dynamic behavior of EPR steering under a relativistic framework

In this section, we assume that there are two parties, Alice and Bob, sharing a pair of particles, which reads as

$$|\varphi\rangle_{AB} = \cos\alpha|00\rangle + \sin\alpha|11\rangle, \quad 0 \leq \alpha \leq \pi/2. \tag{11}$$

It is entangled when $0 < \alpha < \pi/2$. Then, one considers that Alice ($A$) is an inertial observer, and Bob ($B$) with a constant acceleration $a$ is an accelerated observer. As shown in Fig. 1, due to the constant acceleration, Bob travels on a hyperbola constrained in the region I which is causally disconnected from region II. The coordinate transformation between Minkowski coordinates $(t, z)$ and Rindler coordinates $(\tau, \zeta)$ is expressed as [38, 50-53]

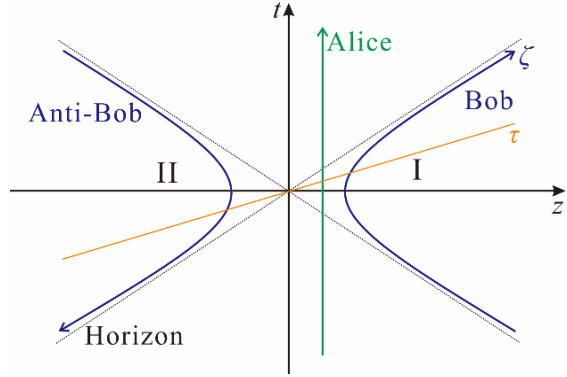

**Fig. 1** (Color online) Rindler space-time diagram: The straight line shows the world line of an inertial observer Alice. An accelerated observer Bob with an acceleration *a* travels on a hyperbolic world line constrained to region I, while a fictitious observer anti-Bob travels on the corresponding hyperbola in region II. Two regions I and II are causally disconnected.

$$at = e^{a\zeta}\sinh(a\tau), \qquad az = e^{a\zeta}\cosh(a\tau). \tag{12}$$

The Rindler coordinates $(\tau, \zeta)$ ranges from $-\infty$ to $\infty$ separately in two regions. This means that each of them admits a separate quantization procedure with the corresponding positive-energy and negative-energy solutions of $\{\psi_{k,s}^{I+}, \psi_{k,s}^{I-}\}$ and $\{\psi_{k,s}^{II+}, \psi_{k,s}^{II-}\}$. Subsequently, the Dirac fields in terms of two sets of the Rindler modes are expanded as [52-54]

$$\psi = \sum_s \int dk \left( c_{k,s}^{I} \psi_{k,s}^{I+} + d_{k,s}^{I\dagger} \psi_{k,s}^{I-} + c_{k,s}^{II} \psi_{k,s}^{II+} + d_{k,s}^{II\dagger} \psi_{k,s}^{II-} \right), \tag{13}$$

where $c_{k,s}^{I}$ and $c_{k,s}^{II}$ are the annihilation operators, $d_{k,s}^{I\dagger}$ and $d_{k,s}^{II\dagger}$ are the creation operators acting on the mode decomposition in regions I and II, respectively. There is a quantum state for Eq. (11) in modes *A* and *B*. One can describe uniformly accelerated Bob to travel on a hyperbola constrained to region I, and Bob has no access to field modes in the causally disconnected region II. And Alice is an inertial observer who stays stationary at a flat region. Consider the Unruh effect in quantum information beyond the single-mode approximation [38]. Let $|0\rangle_U$ and $|1\rangle_U$ be Unruh vacuum state and one-particle state respectively, which reads as [38, 51-53]

$$|0\rangle_U \rightarrow \cos\beta(\cos\beta|0000\rangle - \sin\beta|0011\rangle) + \sin\beta(\cos\beta|1100\rangle - \sin\beta|1111\rangle),$$
$$|1\rangle_U \rightarrow q_R(\cos\beta|1000\rangle - \sin\beta|1011\rangle) + q_L(\cos\beta|0001\rangle + \sin\beta|1101\rangle),$$
(14)

respectively. Here, $\beta = \arccos(1+e^{-2\pi\omega/a})^{-1/2}$ is defined as the acceleration parameter with $\beta \in [0, \pi/4]$, $a \in [0, \infty)$, and $\omega$ is the frequency of Unruh mode. $q_R$ and $q_L$ are complex numbers with $|q_R|^2 + |q_L|^2 = 1$. One utilizes the notation $|mnjk\rangle = |m\rangle_I^+ |n\rangle_{II}^- |j\rangle_I^- |k\rangle_{II}^+$, where the subscripts $\pm$ indicate particle and antiparticle, and the subscripts I and II refer to the regions I and II, respectively.

From now on, for both convenience and simplicity, we restrict our consideration to the cases in which both $q_R$ and $q_L$ are real numbers. In Eq. (14), the single-mode approximation can be established by setting $q_R = 1$, so that the vacuum state and one-particle state can be written as [51]

$$|0\rangle_V \rightarrow \cos\beta|0\rangle_I |0\rangle_{II} + \sin\beta|1\rangle_I |1\rangle_{II},$$
$$|1\rangle_V \rightarrow |1\rangle_I |0\rangle_{II},$$
(15)

respectively.

The initial state is prepared in modes *A* and *B*. Under such a transformation, mode *B* is mapped into two sets of regions for the outside region I and the inside region II, respectively. The complete description of the system involves three modes: mode *A*, described by Alice; mode $B_I$, described by the observer Bob of outside region; and mode $B_{II}$, described by a hypothetical observer anti-Bob confined inside region. Then, by tracing over the state in region II, the reduced density matrix shared by Alice and Bob can be expressed as

$$\rho_{AB_I} = \cos^2\alpha\cos^2\beta|00_I\rangle\langle 00_I| + \cos\alpha\sin\alpha\cos\beta|00_I\rangle\langle 11_I| + \sin^2\alpha|11_I\rangle\langle 11_I|$$
$$+ \cos^2\alpha\sin^2\beta|01_I\rangle\langle 01_I| + \cos\alpha\sin\alpha\cos\beta|11_I\rangle\langle 00_I|.$$
(16)

In Table 1, we give the corresponding each parameter expression of three bipartite mixed states in Bloch decomposition. Here, the corresponding reduced density matrices $\rho_{AB_I}$, $\rho_{AB_{II}}$ and $\rho_{B_I B_{II}}$ can be obtained, by tracing over $B_{II}$, $B_I$ and $A$, respectively. Then, it is straightforward to insert each parameter of the bipartite entangled mixed states (EMSs) $\rho_{AB_I}$ into Eqs. (7) and (8), resulting in the analytical expressions of EPR steering $S^{A \rightarrow B_I}$. In the same way, using Eqs. (9) and (10), one can obtain analytical expressions of EPR steering $S^{B_I \to A}$.

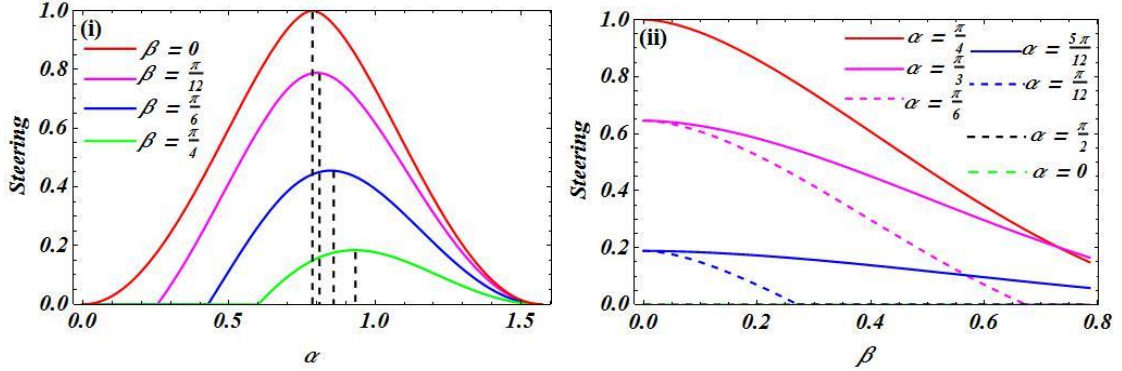

**Fig. 2** (Color online) **(i)** EPR steering $A \to B_I$ as function of state parameter $\alpha$ for different acceleration parameter $\beta$. **(ii)** The EPR steering $A \to B_I$ as function of acceleration parameter for different state parameter.

**Table 1.** The corresponding expressions of each parameter of three different bipartite EMSs in Bloch decomposition are given.

|       | $\rho_{AB_I}$ | $\rho_{AB_{II}}$ | $\rho_{B_I B_{II}}$ |
|-------|---------------|------------------|---------------------|
| $c_1$ | $\sin(2\alpha)\cos\beta$ | $\sin(2\alpha)\sin\beta$ | $\sin(2\beta)\cos^2\alpha$ |
| $c_2$ | $-\sin(2\alpha)\cos\beta$ | $\sin(2\alpha)\sin\beta$ | $-\sin(2\beta)\cos^2\alpha$ |
| $c_3$ | $\cos(2\beta)\cos^2\alpha+\sin^2\alpha$ | $\cos(2\beta)\cos^2\alpha-\sin^2\alpha$ | $\cos(2\alpha)$ |
| $r$   | $\cos(2\alpha)$ | $\cos(2\alpha)$ | $\cos(2\beta)\cos^2\alpha-\sin^2\alpha$ |
| $s$   | $\cos(2\beta)\cos^2\alpha-\sin^2\alpha$ | $\cos(2\beta)\cos^2\alpha+\sin^2\alpha$ | $\cos(2\beta)\cos^2\alpha+\sin^2\alpha$ |

Now, we investigate EPR steering $S^{A \to B_I}$ in the presence of Unruh effect caused by the acceleration of Bob in Fig. 2. One can see that the overall trend of EPR steering decreases with the increase of acceleration parameter $\beta$ in Fig. (ii). The EPR steering $A \to B_I$ maybe disappears if Unruh effect is very stronger. From the Fig. 2 (i), the EPR steering $A \to B_I$ firstly increases and then decreases with the increase of the state parameter $\alpha$ for a fixed $\beta$. It is interesting that the red curve of the EPR steering (without Unruh effect) is symmetrical, while purple, blue and green curves are not. Besides, the value of the state parameter $\alpha$, which corresponds to the position of the EPR steering's maximum, increases as acceleration parameter grows. It turns out that Unruh effect can destroy the symmetry of EPR steering for inertial state or the initial state. In addition, we can obtain that the terminal of all curves for EPR steering are superposition and zero, which means that the corresponding state is a product state at the moment, no entanglement and no EPR steering (see Fig. 2 (i)).

Furthermore, in order to explore the relationship between EPR steering $A \to B_1$ and Bell nonlocality, the Bell Clauser-Horne-Shimony-Holt (CHSH) inequality is introduced. According to the Horodecki criterion [1, 2], $B = 2\sqrt{\max_{i<j}(\mu_i + \mu_j)}$ with $i, j = 1, 2, 3$. The corresponding three eigenvalues $\mu_i$ of $U = T^T T$ are

$$\mu_1 = 4(|\rho_{14}| + |\rho_{23}|)^2, \ \mu_2 = 4(|\rho_{14}| - |\rho_{23}|)^2, \ \mu_3 = (\rho_{11} - \rho_{22} - \rho_{33} + \rho_{44})^2. \tag{17}$$

It is easy to see that $\mu_1$ is always larger than $\mu_2$, and thus the Bell-CHSH inequality maximum violation for X-state is given by [49, 55-57]

$$B(\rho) = 2\max\{B_1, B_2\}, \ B_1 = \sqrt{\mu_1 + \mu_2}, \ B_2 = \sqrt{\mu_1 + \mu_3}, \tag{18}$$

choosing a convenient normalization, we define the Bell nonlocality quantifier as

$$BN := \max\left\{0, \ \frac{B(\rho) - 2}{B_{\max}(\rho) - 2}\right\}. \tag{19}$$

Because $B(\rho) \leq B_{\max}(\rho) = 2\sqrt{2}$ (The bounds of both Tiserlson's and Cirel'son's [55, 56]), one has that $BN \in [0, 1]$. Based on Eqs. (18) and (19), we can obtain the expression of Bell nonlocality for $\rho_{AB_1}$

$$BN = \max\left\{0, \ \frac{B(\rho_{AB_1}) - 2}{2\sqrt{2} - 2}\right\}. \tag{20}$$

with $B(\rho_{AB_1}) = \max\left\{4\sqrt{2}|\cos\alpha\sin\alpha\cos\beta|, \ 2\sqrt{4\cos^2\alpha\cos^2\beta\sin^2\alpha + (\cos^2\alpha\cos(2\beta) + \sin^2\alpha)^2}\right\}$.

As shown in Fig. 3, one can readily find that the EPR steering $S^{A \to B_1}$ and Bell nonlocality decrease with increasing of acceleration parameter, and firstly increase and then decrease with the increase of state parameter. From the Fig. 3 (1), the EPR steering not disappear for $\alpha > 0.6$ whatever the acceleration parameter is, which means that the EPR steering $S^{A \to B_1}$ not experiences a sudden death with increasing of acceleration parameter. However, Bell nonlocality disappears for $\beta \to \pi/4$ in Fig. 3 (2). It means that some steerable states cannot satisfy Bell nonlocality, while the satisfied Bell nonlocality's states are steerable. In addition, EPR steering $S^{A \to B_1}$ and Bell nonlocality are much stronger when the EMSs is a maximally entangled mixed state, that is $\alpha=\pi/4$. If one could use low acceleration to drive Bob, this assure that all EMSs can be employed to realize EPR steering from Alice to Bob and satisfy Bell nonlocality when the state parameter is approximatively greater than 0.6 and less than $\pi/2$.

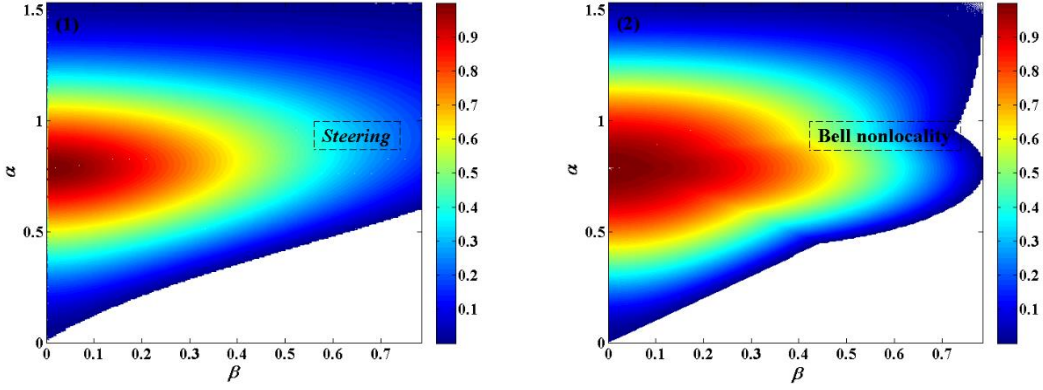

**Fig. 3** (Color online) Contour plot of EPR steering $A \to B_I$ and Bell nonlocality versus acceleration parameter $\beta$ and state parameter $\alpha$ in **(1)** and **(2)**, respectively.

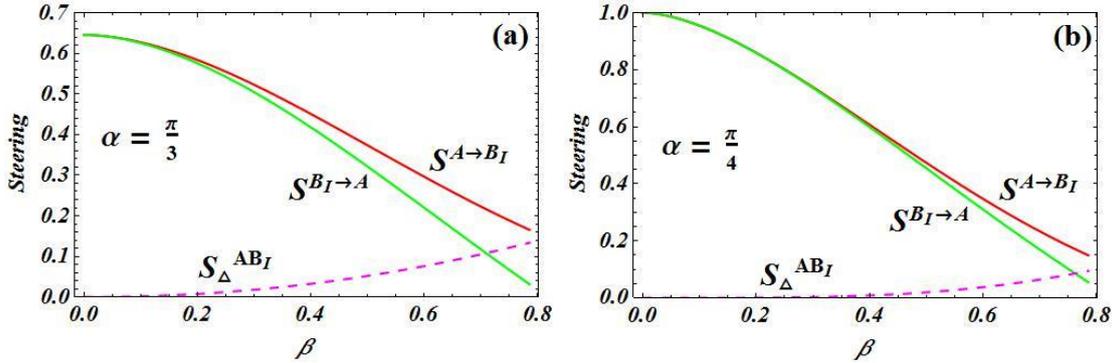

**Fig. 4** (Color online) The EPR steering $S^{A \to B_I}$ (red solid line), $S^{B_I \to A}$ (green solid line), and the EPR steering asymmetry $S_\Delta^{AB_I}$ (purple dashed line) between Alice and Bob as a function of the acceleration parameter, when **(a)** $\alpha=\pi/3$ and **(b)** $\alpha=\pi/4$.

Moreover, to check the degree of steerability asymmetric under the relativistic framework, we define EPR steering asymmetry as $S_\Delta^{AB_I} = |S^{A \to B_I} - S^{B_I \to A}|$. Then, we plot the EPR steering $S^{A \to B_I}$, $S^{B_I \to A}$ as well as EPR steering asymmetry $S_\Delta^{AB_I}$ as a function of $\beta$ for a fixed state parameter $\alpha=\pi/3$ or $\alpha=\pi/4$ in Fig. 4. The EPR steering $S^{A \to B_I}$ and $S^{B_I \to A}$ are a monotonically decreasing function of acceleration parameter, which means that decoherence introduced by Unruh effect can cause the degradation of EPR steering shared between an inertial observer and an accelerated one. It is indicated that EPR steering $S^{B_I \to A}$ is weaker than EPR steering $S^{A \to B_I}$ and also avoids sudden death with the increase of acceleration parameter, which denotes that EPR steering from non-inertial part to the inertial one is more difficult than EPR steering from inertial part to the non-inertial one. That is to say, EPR steering in a bipartite quantum system can be seen as a better resource for implementing quantum information processing. Furthermore, one can obtain that $S^{A \to B_I} \neq S^{B_I \to A}$ when the acceleration parameter is greater than a fixed value, after that, EPR steering has an asymmetry property. The EPR steering asymmetry increases with enlarging intensity of Unruh effect, which means that the Unruh effect can enhance EPR steering asymmetry, it may be caused by Unruh effect destroyed the steerability of the state.

For the sake of better understanding the interaction between state parameter and Unruh effect in the generation of EPR steering asymmetry, we plot EPR steering asymmetry $S_\Delta^{AB_I}$ as functions of acceleration parameter and state parameter in Fig. 5. One can obtain that EPR steering asymmetry monotonically increases with increasing acceleration parameter, which means that the resources shared in the EMSs play a primary role in EPR steering as well. Furthermore, EPR steering asymmetry amplifies its value with increasing acceleration parameter $\beta$, and that the maximum of EPR steering asymmetry does not exceed $0.2\ln 2$. The EPR steering asymmetry can be generated when the state parameter is approximately greater than $0.6$ and less than $\pi/2$.

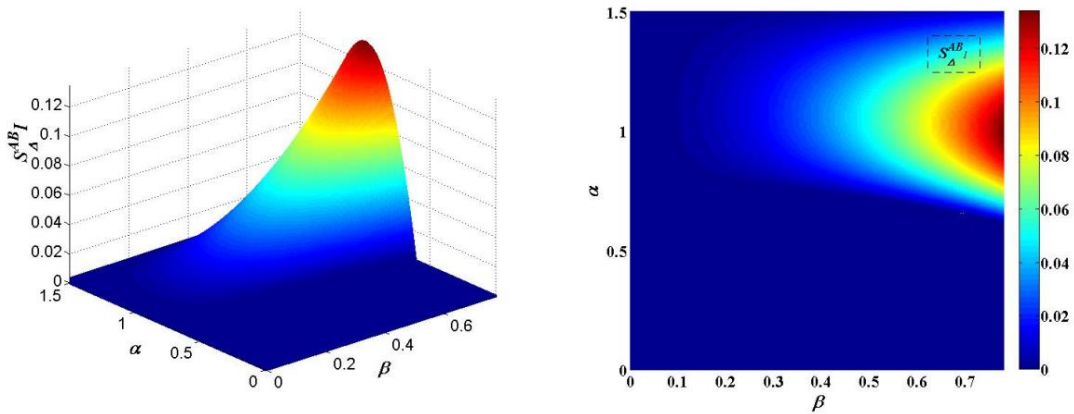

**Fig. 5** (Color online) (left) EPR steering asymmetry $S_\Delta^{AB_I}$ as functions of acceleration parameter and state parameter. (right) Contour plot of EPR steering asymmetry $S_\Delta^{AB_I}$ versus acceleration parameter and state parameter.

## 4. EPR steering redistribution

In order to better unveil why the physical accessible EPR steering can be reduced with the growing acceleration parameter, an open question has been put forward: whether the lost EPR steering is destroyed or transferred to anywhere? To solve this problem, we calculate EPR steering of other possible bipartite states which are physically inaccessible. We guess that the disappeared EPR steering is distributed to the physical inaccessible region. We here still consider the case of $q_R = 1$ in Eq. (14), and then one can obtain the corresponding reduced density matrices of partition $AB_{II}$ and $B_I B_{II}$, by tracing over $B_I$ and $A$, respectively.

$$\rho_{AB_{II}} = \cos^2\alpha\cos^2\beta |00_{II}\rangle\langle 00_{II}| + \cos\alpha\sin\alpha\cos\beta |01_{II}\rangle\langle 10_{II}| + \sin^2\alpha |10_{II}\rangle\langle 10_{II}| \\ + \cos^2\alpha\sin^2\beta |01_{II}\rangle\langle 01_{II}| + \cos\alpha\sin\alpha\cos\beta |10_{II}\rangle\langle 01_{II}|, \qquad (21)$$

and

$$\rho_{B_I B_{II}} = \cos^2\alpha\cos^2\beta |0_I 0_{II}\rangle\langle 0_I 0_{II}| + \sin\beta\cos\beta\cos^2\alpha |0_I 0_{II}\rangle\langle 1_I 1_{II}| + \sin^2\alpha |1_I 0_{II}\rangle\langle 1_I 0_{II}| \\ + \sin\beta\cos\beta\cos^2\alpha |1_I 1_{II}\rangle\langle 0_I 0_{II}| + \cos^2\alpha\sin^2\beta |1_I 1_{II}\rangle\langle 1_I 1_{II}|, \qquad (22)$$

Afterwards, via Eqs. (7), (8) and Table. 1, we can obtain the corresponding analytical expressions of EPR steering $S^{A \to B_{II}}$ and $S^{B_I \to B_{II}}$, respectively.

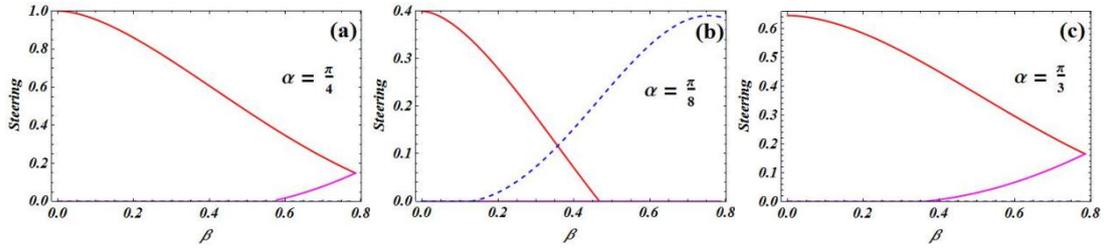

**Fig. 6** (Color online) EPR steering redistribution: EPR steering of three different bipartite states as function of acceleration parameter $\beta$, when **(a)** $\alpha = \pi/4$, **(b)** $\alpha = \pi/8$, **(c)** $\alpha = \pi/3$. The red solid line, purple solid line and blue dotted line denote EPR steering $S^{A \to B_I}$, $S^{A \to B_{II}}$ and $S^{B_I \to B_{II}}$, respectively.

As shown in Fig. 6, the redistributions of EPR steering show how the Unruh effect changes all nonzero bipartite EPR steering. It demonstrates that modes $A$ and $B_I$ remain maximally steering, while the values of other EPR steering are close to zero when the Unruh effect is feeble. As the intensity of Unruh effect grows, the physical accessible EPR steering decreases, while the physical inaccessible EPR steering increases until that the acceleration parameter is big enough. In other words, the physical accessible EPR steering and physical inaccessible EPR steering are approximately trading off and taking turns. Consequently, we get a conclusion that the disappeared physical accessible EPR steering from Alice to Bob is distributed to the physical inaccessible EPR steering (from Alice to anti-Bob or from Bob to anti-Bob). It is also worth mentioning that EPR steering from Alice to Bob and from Alice to anti-Bob are equivalent, when the acceleration parameter is infinite $\beta \to \pi/4$. Additionally, the EPR steering from Alice to anti-Bob and from Bob to anti-Bob cannot synchronously appear, arising from the monogamy relationship of EPR steering: two systems *A* and *C* cannot both steer the third system *B*. This is quite different from the behavior of quantum discord and entanglement redistribution under the relativistic setting [48]. We can also obtain that the EPR steering from Alice to anti-Bob appears and another EPR steering from Bob to anti-Bob disappears when the state parameter is approximately larger than 0.6 less than $\pi/2$ in Fig. 7.

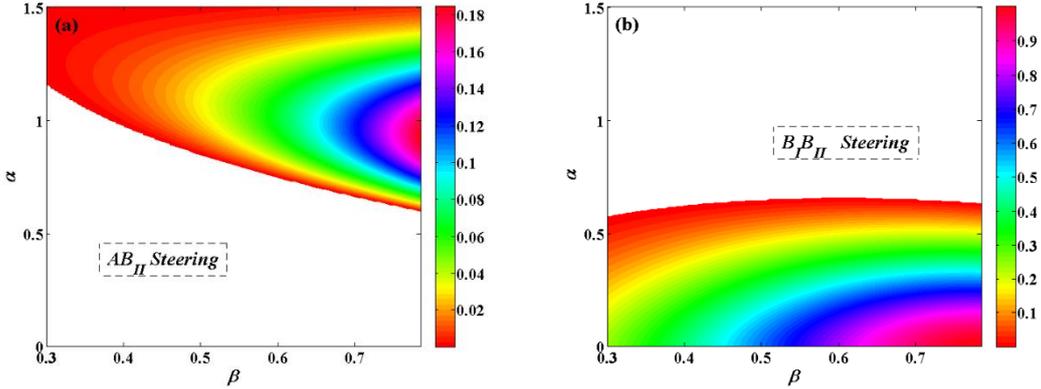

**Fig. 7** (Color online) Contour plot of EPR steering versus acceleration parameter $\beta$ and state parameter $\alpha$. The (a) and (b) denote EPR steering $S^{A \to B_{II}}$ and $S^{B_I \to B_{II}}$ of the states $\rho_{AB_{II}}$ and $\rho_{B_I B_{II}}$, respectively.

## 5. Summary

To summarize, we observed the dynamic behavior of EPR steering and discussed EPR steering redistribution under a relativistic framework. The results indicate that the maximal entangled mixed state is maximally steerable state and not all EMSs are steerable. In other words, if an entangled state shared by Alice and Bob is steerable, after Bob suffers from Unruh noise, and the steerable state may become unsteerable. Besides, EPR steering from Alice to Bob experiences a sudden death with increasing acceleration parameter when $\alpha$ is approximately less than 0.6. That is, the Unruh effect can destroy EPR steering from Alice to Bob. However, EPR steering from Bob to anti-Bob and from Alice to anti-Bob experience a sudden birth with increasing acceleration parameter. In addition, EPR steering has also an asymmetry property under the relativistic framework, and the EPR steering asymmetry increases with increasing intensity of Unruh effect, which displays that the Unruh effect can enhance EPR steering asymmetry, and its effect can weaken the steerability of the states and destroyed the symmetry of the states.

Furthermore, there exists a trade-off: by increasing the intensity of Bob's acceleration, we obtained that the physical inaccessible EPR steering increases, but the physical accessible EPR steering decreases. That is, the reduced physical accessible EPR steering is distributed to physical inaccessible EPR steering. Unlike quantum discord and entanglement [48], EPR steering of the states $\rho_{AB_{II}}$ and $\rho_{B_{I}B_{II}}$ cannot synchronously appear. Namely, the monogamy relationship of EPR steering [37] is also hold under the relativistic setting. This is also quite different from the behavior of entanglement redistribution. Consequently, we believe that EPR steering could be used as one of important resources within the long-distance quantum secure communication under the relativistic framework. What's more, some interesting open questions are unresolved including: how to establish or quantify the relationship between EPR steering and Bell nonlocality, and how the dynamic behavior of EPR steering is under other relativistic frameworks, such as Schwarzschild space-time [58] and Garfinkle-Horowitz-Strominger dilation space-time [59].

## Acknowledgments

This work was supported by the National Natural Science Foundation of China under Grant Nos. 11575001, 61601002 and 11605028, Anhui Provincial Natural Science Foundation (Grant No. 1508085QF139), and the open fund from CAS Key Laboratory of Quantum Information, University of Science and Technology of China (Grant No. KQI201701).